\documentclass[a4paper,fleqn,usenatbib]{mnras}
\usepackage{graphicx}	
\usepackage{amsmath}	
\usepackage{mathtools}
\usepackage{amssymb}	
\usepackage{comment,orcidlink}

\def\be{\begin{equation}}
\def\ee{\end{equation}}

\def\del#1{{}}

\newcommand\mj{{\,{\rm M}_{\rm J}}}

\newcommand\MSunPerYear{~${\rm M_{\odot}}$~yr$^{-1}$\,}

\newcommand{\bref}[1]{{ #1}}

\title[]{Dust growth and planet formation by disc fragmentation}

\author[Lee et al.]{Hans Lee$^{1 \orcidlink{0009-0006-6924-8962}}$, Sergei Nayakshin$^{1 \orcidlink{0000-0002-6166-2206}}$\thanks{sergei.nayakshin@le.ac.uk} and Richard A. Booth$^{2 \orcidlink{0000-0002-0364-937X}}$\\
$^{1}$School of Physics and Astronomy, University of
  Leicester, Leicester, LE1 7RH, UK \\
  $^2$School of Physics and Astronomy, University of Leeds, Leeds LS2 9JT, UK \\
}

\date{Accepted XXX. Received YYY; in original form ZZZ}

\pubyear{2025}

\begin{document}
\label{firstpage}
\pagerange{\pageref{firstpage}--\pageref{lastpage}}

\maketitle

\begin{abstract}
It is often argued that gravitational instability of realistic protoplanetary discs is only possible at distances larger than $\sim 50$~au from the central star, requiring high disc masses and accretion rates, and that therefore disc fragmentation results in the production of brown dwarfs rather than gas giant planets. However, the effects of dust growth on opacity can be very significant but have not been taken into account systematically in the models of fragmenting discs. We employ dust opacity that depends on both temperature and maximum grain size to evaluate analytically  the properties of a critically fragmenting protoplanetary disc. We find that dust growth may promote disc fragmentation at disc radii as small as $\sim 30$~au. As a result, the critical disc masses and accretion rates are smaller, and the initial fragment masses are in the gas giant planet mass regime. While this suggests that formation of gas giant planets by disc fragmentation may be more likely than usually believed, we caution that numerical models of the process are needed to evaluate the effects not taken into account here, e.g., dust grain mobility and fragment evolution after disc fragmentation.

\end{abstract}


\section{Introduction}

Planet formation by Gravitational Instability \citep[GI; see reviews by][]{HelledEtal13a,KratterL16} posits that young massive protoplanetary discs fragment gravitationally into clumps \citep[e.g.][]{Kuiper51,Boss98}, which then evolve further to form planets \citep{Nayakshin_Review} or brown dwarfs \citep{SW09b}.

It is universally accepted that to fragment, discs must be massive and cool \citep{Toomre64}, and also be able to cool rapidly \citep{Gammie01,MeruBate10b,DengEtal17}; otherwise, a self-regulated gravito-turbulent state develops in which disc density perturbations in the disc constantly appear and dissolve away \citep{Rice05,CossinsEtal09,Paardekooper12a}. The disc fragmentation conditions are believed to be satisfied only at large distances, i.e., $R\gtrsim 50$~au \citep{Rafikov05, ClarkeLodato09}. Giant planets discovered orbiting their stars on wide orbits \citep[e.g.][]{MaroisEtal08,MaroisEtal10,Vigan21-SHINE, Blunt23} may have formed via GI.

However, the potential of GI to make planets rather than brown dwarfs is frequently questioned since both analytical estimates \citep{KratterEtal10,ForganRice11} and most numerical simulations \citep{SW08,ZhuEtal12a,Xu24} indicate initial fragment masses $M_{\rm frag}\gtrsim 10\mj$, although note that some numerical studies do suggest that sub-Jovian mass gaseous objects can be formed by GI \citep[e.g.][]{BoleyEtal10,Kubli-GI-MHD-23}.

Turbulence in protoplanetary discs may result in very rapid grain growth \citep{Weidenshilling84,DD05}. A number of authors \citep[e.g.][]{RiceEtal04,BoleyDurisen10,BoothClarke16,Vorobyov-Elbakyan-18,Vorobyov-Elbakyan-19} pointed out that dust in self-gravitating discs can grow to significant sizes, $a_{\rm max} \sim 1-100$~cm in the first $\sim 0.1$ Myr \citep[e.g.,][]{Molyrova_21}. Furthermore, focused in spiral density arms \citep{RiceEtal04}, large grains can become self-gravitating and collapse into massive solid cores \citep{GibbonsEtal12,GibbonsEtal14,Baehr-23-GI-dust,Longarini23a,Longarini23b,Rowther24}, which could later turn into gas giant planets via gas accretion as in the Core Accretion scenario \citep[CA,][]{HelledEtal13a} or the Pebble Accretion Scenario \citep{Baehr-23-GI-dust}.

In this Letter, we draw attention to the complementary and potentially stronger effects of dust growth on planet formation by GI due to opacity effects. Although dust makes up approximately 1\% of the mass of protoplanetary discs, it contributes $\gtrsim 99.9$\% of the total opacity \citep[e.g. cf. Fig. 5 in][]{Malygin14-opacity}.

Dust opacity depends significantly on the distribution of particle sizes \citep[e.g.][]{Draine-Lee-84,Woitke16-DIANA}. For a collection of spherical dust particles with uniform size $a$ and fixed total mass, its total geometric cross-sectional area scales as $\propto a^{-1}$, and therefore, the Rosseland mean opacity usually decreases with grain growth \citep[although the behavior of frequency-dependent opacity is more nuanced, e.g., see the ``opacity cliff" in Fig. 1 in][]{Rosotti19-Opacity-Cliff}.

\begin{figure}
\includegraphics[width=\columnwidth]{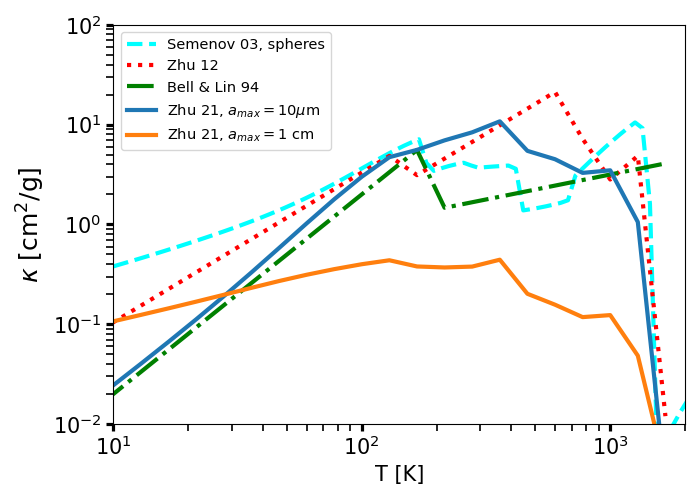}
\caption{Comparison between dust opacities of \protect\cite{Bell94}, \protect\cite{SemenovEtal03}, \protect\cite{ZhuEtal12a} at $\rho = 10^{-10}$ g cm$^{-3}$, and \protect\cite{Zhu+21} at $a_\text{max} = 10 \mu$m and $a_\text{max} = 1$ cm respectively.}
\label{fig:past_opacity}
\end{figure}

Figure \ref{fig:past_opacity} shows several commonly used Rosseland mean opacities for dusty discs. Except for opacities of \cite{Zhu+21}, all of these consider only relatively small dust grains, with maximum grain sizes $a_{\rm max}\lesssim 10\mu$m. \cite{Zhu+21}, focusing on the physics of planetary atmospheres, calculated dust opacities with grain sizes up to 10 cm. Fig. \ref{fig:past_opacity} shows clearly that while dust opacity depends significantly on grain physics, it is most sensitive to $a_{\rm max}$, and can drop by one-two orders of magnitude in an important range of disc temperatures if grains can grow to sizes larger than $\sim 1$~cm. 

Below we present an analytic dust opacity fit for dust opacity computed for a wider range of dust sizes than in \cite{Zhu+21}, and use it to re-evaluate disc fragmentation conditions and initial fragment masses.

\section{Dust opacity}\label{sec:opac}

\subsection{An analytic fit to opacity curves}

We use the DIANA standard opacities described in \cite{Woitke16-DIANA}, utilising the OpacityTool Fortran package \citep[][]{Toon81, Dorschner95, Zubko96, Min05} to produce tables of dust absorption coefficient $\kappa$ for a range of wavelengths and $a_{\rm max}$ values. We assume that the number density of dust grains follows a power law: $n(a)da \propto a^{-p}$, where $p = 3.5$, from the minimum $a_{\rm min} = 0.05\mu$m up to the maximum dust size $a_{\rm max}$. These opacity tables are calculated for the wavelength range $0.05 \mu$m $\leq \lambda \leq 9000 \mu$m, which covers the blackbody spectrum for the temperature range we are interested in sufficiently well.

We produce 17 of these opacity tables, for values of $a_\text{max}$ evenly spaced on the logarithmic scale from $1 \mu$m to $100$ m. Since our focus is on the total cooling rate via radiation in optically thick discs, we calculate the Rosseland mean opacity $\kappa_\text{DIANA}(T)$ for each opacity table, which are functions of temperature.

For a given $a_\text{max}$, we then fit the DIANA opacity curves to the following piecewise power law function:

\begin{equation}
    \kappa(a_{\rm max}, T) = \begin{dcases}
        \kappa_0\Bigl(\dfrac{T}{155 \text{ K}}\Bigr)^{p_l} & T \leq 155 \text{ K} \\
        \kappa_0 & 155 \text{ K} \leq T \leq 353 \text{ K} \\
        \kappa_0\Bigl(\dfrac{T}{353 \text{ K}}\Bigr)^{p_h} & T \geq 353 \text{ K}
    \end{dcases}
    \label{analytic_fit}
\end{equation}

where $\kappa_0$, $p_l$, and $p_h$ vary with $a_\text{max}$. \bref{The  transition temperatures are chosen empirically according to the shape of the opacity curves, and are not of particular physical significance.}

The values of these parameters for each $a_\text{max}$ are listed in Table \ref{tab:params}. Using this, one can obtain intermediate values by interpolating $p_l$, $p_h$, and $\log \kappa_0$ linearly with respect to $\log a_{\rm max}$.

\begin{table}
    \centering
    \begin{tabular}{c|ccc}
         $a_\text{max}$ [$\mu$m] & \bref{$\kappa_0$} & $p_l$ & $p_h$ \\
         \hline
         $1.0$ & 1680 & 1.55 & 1.05 \\
         $3.16$ & 2160 & 1.64 & 0.867 \\
         $10.0$ & 2490 & 1.64 & 0.516 \\
         $31.6$ & 2040 & 1.40 & 0.311 \\
         $1.0 \times 10^2$ & 1090 & 0.905 & 0.369 \\
         $3.16 \times 10^2$ & 564 & 0.640 & 0.432 \\
         $1.0 \times 10^3$ & 297 & 0.515 & 0.442 \\
         $3.16 \times 10^3$ & 173 & - & - \\
         $1.0 \times 10^4$ & 98.5 & - & - \\
         $3.16 \times 10^4$ & 55.8 & - & - \\
         $1.0 \times 10^5$ & 31.4 & - & - \\
         $3.16 \times 10^5$ & 17.7 & - & - \\
         $1.0 \times 10^6$ & 9.92 & - & - \\
         $3.16 \times 10^6$ & 5.56 & - & - \\
         $1.0 \times 10^7$ & 3.12 & - & - \\
         $3.16 \times 10^7$ & 1.75 & - & - \\
         $1.0 \times 10^8$ & 0.982 & - & - \\
    \end{tabular}
    \caption{Fit parameters \bref{$\kappa_0$}, $p_l$, and $p_h$ for the range of $a_\text{max}$ values computed. For $a_\text{max} \geq 10^3 \mu$m, $p_l$ and $p_h$ are constant. Note that this function gives the opacity for dust only ($z=1$), and should be multiplied by the metallicity of the system.}
    \label{tab:params}
\end{table}

For $a_\text{max} \geq 1$mm, the shape of the curve is unchanged on a log-log plot, therefore $p_l$ and $p_h$ are fixed. When $a_\text{max}$ increases by a factor of 10, $\kappa_0$ increases by a factor of $\approx 3$. 

\begin{figure}
\includegraphics[width=\columnwidth]{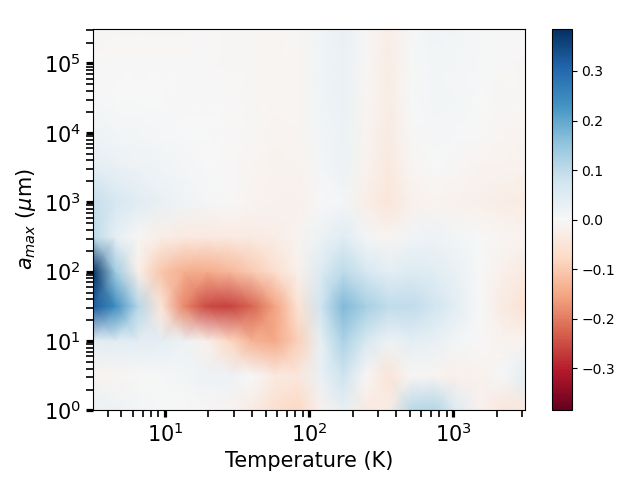}
\caption{Logarithm of the opacity ratio between the analytic function and the DIANA opacity curves.}
\label{fig:fit_heatmap}
\end{figure}

We characterise the goodness of fit by calculating the ratio between our function $\kappa(\kappa_0, p_l, p_h)$ and the DIANA opacities $\kappa_\text{DIANA}$ by calculating $\log_{10}\Bigl(\dfrac{\kappa(\kappa_0, p_l, p_h)}{\kappa_\text{DIANA}}\Bigr)$, a function of $T$ and $a_\text{max}$, which is shown in figure \ref{fig:fit_heatmap}.

The deviation reaches up to a factor of $10^{0.385} = 2.43$ at $a_\text{max} \sim 100 \mu$m and $T \sim 3$ K, and $10^{0.262} = 1.83$ at $a_\text{max} \sim 30 \mu$m and $T \sim 25$ K. However, figure \ref{fig:past_opacity} demonstrates that there exists a range of reasonable assumptions about dust composition and physics which results in yet larger uncertainties than the errors of the semi-analytic opacity fit. Therefore, we consider eq. \ref{analytic_fit} to be a sufficiently good fit to the DIANA $a_{\rm max}$-dependent opacity curves.

\subsection{Sublimation}
\begin{figure}
\includegraphics[width=\columnwidth]{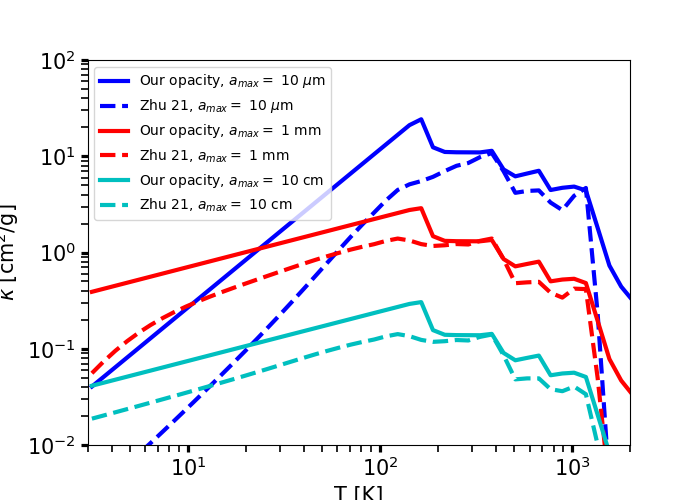}
\caption{Comparison between Rosseland mean opacities in \protect\cite{Zhu+21} (at $\rho = 10^{-10}$ g cm$^{-3}$) and this paper.}
\label{fig:compare}
\end{figure}

Since the DIANA standard opacity calculation does not account for dust sublimation, we estimate how much opacity remains above each sublimation point using the results from \cite{PollackEtal94}, scaled according to the material composition used in DIANA (\cite{Min11}). We first consider the sublimation of three species: water ice, organics, and troilite. We represent the effects of this by multiplying the opacity function by $\bigl(f_i + (1-f_i)\exp(\dfrac{T-T_i}{10\,{\rm K}})\bigr)$ when $T > T_i$ for all species $i$. $T_i$ is the sublimation temperature for each species considered, and $f_i$ is the remaining opacity fraction after sublimation. The values used are shown in table \ref{tab:sublimation}, \bref{and as shown in fig. \ref{fig:compare}, there is a notable sharp decrease in opacity at each sublimation temperature.}

For the evaporation of the most refractory grains, we follow \cite{Kuiper10}, multiplying the dust opacity by $0.5 - \dfrac{1}{\pi} \arctan \Bigl(\dfrac{T-T_\text{evap}}{100\,{\rm K}} \Bigr)$, where $T_\text{evap} = 2000(\dfrac{\rho}{1\,{\rm g\,cm}^{-3}})^{0.0195}\, {\rm K}$, where $\rho$ is the gas density.

\begin{table}
    \centering
    \begin{tabular}{c|cc}
         Material & $T_i$ [K] & $f_i$ \\
         \hline
         Water ice & 166 & 0.455 \\
         Organics & 425 & 0.472 \\
         Troilite & 680 & 0.593 \\
    \end{tabular}
    \caption{$T_i$ and $f_i$ used to account for sublimation in our opacity fit.}
    \label{tab:sublimation}
\end{table}

The resulting opacity function is shown in figure \ref{fig:compare}, alongside the dust size-dependent opacities from \cite{Zhu+21}. \bref{The difference between the curves is mainly due to the assumption we make about the dust composition. For example, the water ice mass fraction we use is about 2 times that of the one in \cite{Zhu+21}. This can be seen in fig. \ref{fig:compare}, where our opacity curve has a greater drop at the water ice sublimation temperature.} Since our opacity is larger at all temperatures and $a_{\rm max}$, we opt to use it in the rest of the paper. This approach is conservative with respect to opacity uncertainties as it leads to higher disc and fragment masses at disc fragmentation, which is the main focus of our paper.

\section{Disc fragmentation with dust growth}\label{sec:mfrag}

In this section, we use the opacity function derived in \S \ref{sec:opac} to calculate the properties of a critically fragmenting disc.

\subsection{Critically fragmenting discs}

Pseudo-viscous disc models is a convenient framework to study the disc fragmentation boundary analytically \citep[e.g.][]{Rafikov05,Levin07,Clarke09}, with their results being consistent with numerical simulations within a factor of order unity \citep[e.g. fig. 2 in][]{ZhuEtal12a}. In this approach, a marginally stable self-gravitating protoplanetary disc has the \cite{Toomre64} parameter $Q=\dfrac{c_s \Omega}{\pi G \Sigma} \simeq 1$, which gives the sound speed $c_s = \dfrac{\pi G \Sigma}{\Omega}$, where $\Omega$ is the orbital angular velocity. The disc mid-plane temperature is then

\begin{equation}
    T = \dfrac{2m_p c_s^2}{k_B} = \dfrac{2m_p}{k_B} \Bigl(\dfrac{\pi G \Sigma}{\Omega}\Bigr)^2.
    \label{eq:temp}
\end{equation}

Further, in marginally stable discs, heating generated by gravitoturbulence is balanced by radiative cooling \citep{Gammie01}. \cite{Xu24} show that the disc's radiative cooling rate is well approximated in both optically thick and optically thin regimes by

\begin{equation}
    F_{\rm cool} = 8\sigma_{\rm B} (T^4 - T_\text{irr}^4) \dfrac{\tau}{(1 + (0.875 \tau^2)^{0.45})^{1/0.45}},
    \label{eq:rad_flux}
\end{equation}

where the optical depth of the disc $\tau = \kappa \Sigma/2$, and the disc temperature due to stellar irradiation heating $T_\text{irr} = \Bigl(\dfrac{0.1 L_*}{4\pi r^2 \sigma_B}\Bigr)^{1/4}$ \citep{ZhuEtal12a}.
\bref{The constant factor of 0.1 is a simple estimate of non-local disc irradiation effects, which depend on the disc global $H/R$ profile and properties of the dust at the height of a few $H$ above the disc midplane \citep[e.g.,][]{CG97}. These effects, including possible disc self-shadowing \citep[e.g.,][]{Bell-99-FUOR-shadow}, cannot be computed in a local disc stability calculation. Nonetheless, our choice of 0.1 is consistent with the flaring angle in our calculations within a factor of order unity. Supplementary materials provide examples of how our results vary depending on the choice of $L_*$, which is equivalent to varying the irradiation angle.} 

On the other hand, the heating rate due to gravitoturbulence with the corresponding viscosity parameter $\alpha$ is given by
\begin{equation}
    F_{\rm heat} = \dfrac{3}{8\pi}\Omega^2 \dot M = \dfrac{9}{8}\alpha \Omega c_s^2 \Sigma = \dfrac{9}{8}\dfrac{\alpha (\pi G)^2 \Sigma^3}{\Omega}.
    \label{eq:accr_flux}
\end{equation}

Setting $F_{\rm cool} = F_{\rm heat}$ for a self-regulating disc, one obtains  an equation consisting only of $T$, $\Omega$, $\alpha$, and physical constants. This equation can be solved numerically to find the critical mid-plane temperature $T_\text{crit}$ of the disc as a function of $\Omega$ for a fixed $\alpha$ \citep{Levin07}. Since we are interested in disc fragmentation, we set $\alpha = \alpha_\text{crit} \simeq 0.06$, above which the disc fragments \citep{Rice05}. Note that, in the language of self-gravitating discs in thermal balance between turbulent heating and radiative cooling, this choice of $\alpha_{\rm c}$ corresponds to the ``$\beta$" cooling parameter of $\beta = 4/[9\gamma(\gamma-1)\alpha_{\rm c}] \approx 7$ \citep[cf. eq. 20 in][]{Gammie01} for $\gamma=5/3$ \citep[as appropriate for the cold regions of the disc;][]{BoleyEtal07}.

From that, we calculate the corresponding disc scale height, $H_\text{crit} = 0.69 \dfrac{c_s}{\Omega}$ \citep[][eq. 11 with $Q=1$]{KratterL16}, surface density, $\Sigma_\text{crit}$ (cf. equation \ref{eq:temp}), and thus arrive at the critical accretion rate above which the disc should fragment \citep[e.g.,][]{Clarke09,ZhuEtal12a}, as well as the initial fragment mass, which we take to be $M_\text{frag} = 57 \Sigma_\text{crit} H_\text{crit}^2$ \citep{Xu24}. While the factor in front of $\Sigma_\text{crit} H_\text{crit}^2$ differs between different studies by an order of magnitude \citep[see][]{KratterL16}, here we follow the  larger value derived by \cite{Xu24} in their radiation hydrodynamics simulations, which leads to a more conservative estimate of the initial fragment masses.

The maximum grain size in a marginally stable disc depends on many parameters of the problem and is a non-local quantity due to dust dynamics being a function of grain size \citep[e.g.,][]{RiceEtal04,BoleyDurisen10,BoothClarke16}. Below, we  consider four representative scenarios to evaluate the effects of dust growth on disc fragmentation. Two of the cases assume a fixed $a_{\rm max}$ independent of disc radius, $R$, $a_{\rm max} = 10 \,\mu$m and $a_{\rm max} = 1$ cm, respectively, roughly indicating the ``no growth" and the significant grain growth limits.

For the third case, we estimate dust growth limited by the grain fragmentation velocity, which we take to be $v_{\rm frag} = 10$ m s$^{-1}$. We calculate the maximum dust size as $a_{\rm max} = \dfrac{2 \Sigma}{\pi \alpha \rho_s} \dfrac{v_{\rm frag}^2}{c_s^2}$ \citep{Birnstiel09}, where $\rho_s$ is the material density of solid dust, which we set $3$ g cm$^{-3}$ here. This model assumes that similarly sized grain collisions dominate both grain growth and fragmentation, and occur with relative velocity $\Delta v \sim \sqrt{\alpha \,\rm St} \, c_s$. Other mechanisms of grain growth such as sweep-up of small grains by larger grains \citep{Xu23-DustGrowth} may allow for growth significantly beyond this barrier, but we do not consider these effects here.

Finally, we use $\text{St} = 0.1$ as an optimistic upper limit to the grain sizes for our fourth case because previous work has shown that self-gravity produces particle motions that are correlated when $\text{St} < 1$ \citep[e.g.][]{BoothClarke16}. As a result, collisions may occur at lower velocities than those estimated by the \citet{Birnstiel09} formulae, unless the large-scale motions associated with the spirals can be efficiently converted into small-scale turbulence. While small-scale motion has been identified in high-resolution 3D simulations \citep{RiolsLatter2017,BoothClarke2019}, its effect on the dust growth has not been quantified. Instead, we estimate the most optimistic case by assuming that the only source of collisions between dust grains is due to the differential motion created by the spiral-induced pressure gradients. Following the arguments laid out in \citet{BoothClarke16}, this produces $\text{St} \lesssim 0.1$ for our parameters. The maximum dust size is then found via $a_{\rm max} = 2 \Sigma \text{St}/(\pi \rho_{\rm s})$.

\subsection{Results}

\begin{figure}
\includegraphics[width=\columnwidth]{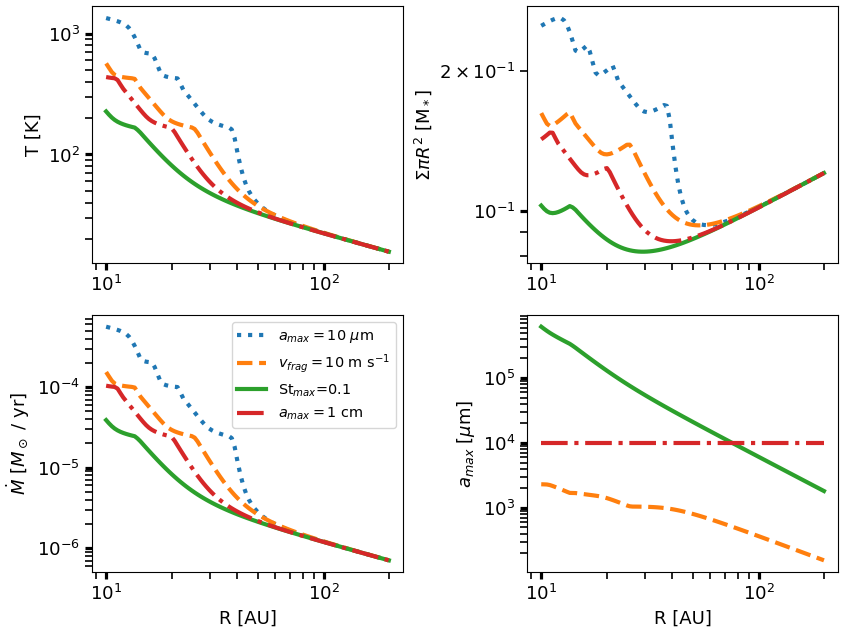}
\caption{Plotted against $R$: $T_\text{crit}$ (upper left), $\Sigma_\text{crit} \pi R^2$ (upper right), $\dot M$ (lower left), $a_{\rm max}$ (lower right). The cases are $a_\text{max} = 10 \mu$m (blue, dotted), $v_{\rm frag} = 10$ m s$^{-1}$ (orange, dashed), $\text{St}_\text{max} = 0.1$ (green, solid), and $a_\text{max} = 1$ cm (red, dash-dotted). We choose to rescale the critical surface density plot by $\pi R^2$ to better illustrate the difference between the four cases.}
\label{fig:critdisc}
\end{figure}

\begin{figure}
\includegraphics[width=\columnwidth]{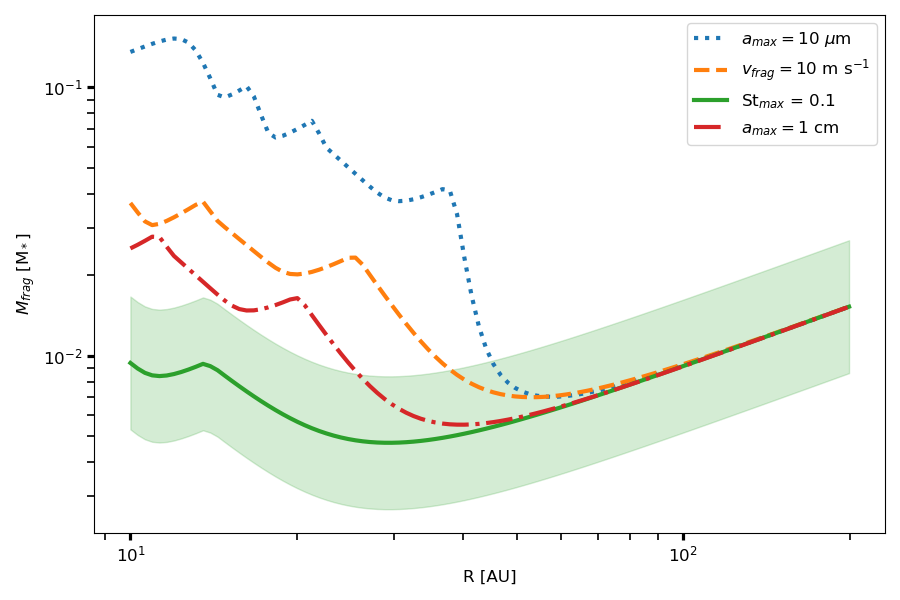}
\caption{$M_\text{frag}$ plotted against $R$, for $a_\text{max} = 10 \mu$m (blue, dotted), $v_{\rm frag} = 10$ m s$^{-1}$ (orange, dashed), $\text{St}_\text{max} = 0.1$ (green, solid), and $a_\text{max} = 1$ cm (red, dash-dotted). $1\sigma$ uncertainty is shown around the $\text{St}_\text{max} = 0.1$ case \citep{Xu24}.}
\label{fig:mfrag}
\end{figure}

Figure \ref{fig:critdisc} shows various properties of the critically fragmenting disc as functions of $R$, demonstrating that dust growth can lead to fragmentation in colder, less massive, and geometrically thinner discs. We observe that up to $R \sim 40$ au, when dust grains are allowed to grow from $a_{\rm max} = 10 \mu$m to $\text{St}_\text{max} = 0.1$, $T_\text{crit}$ decreases by a factor of $5$ to $10$, and $\Sigma_\text{crit}$ decreases by a factor of $\approx 3$.

In the plot of $a_{\rm max}$, we see that for the radius range we consider, $\text{St}_\text{max} = 0.1$ is strictly above $a_\text{max} = 1$ cm, which is then strictly above $v_{\rm frag} = 10$ m s$^{-1}$. 

Dust growth also decreases the critical accretion rate $\dot M_\text{crit}$ for fragmentation. From equation \ref{eq:accr_flux}, $\dot M = 3 \pi \alpha c_s^2 \Sigma \Omega^{-1}$, implying $\dot M_\text{crit} \propto T_\text{crit} \Sigma_\text{crit}$. Since $T_\text{crit}$ and $\Sigma_\text{crit}$ decrease when dust growth is present, $\dot M_\text{crit}$ decreases accordingly. Hydrodynamical simulations of disc formation from protostellar cloud collapse show that accretion rates on the order of $\dot M \sim 10^{-5}$ \MSunPerYear can be expected \citep{VB15}. In the ISM-like case, we find that $\dot M > \dot M_{\rm crit}$ only for $R > 40$ au, while this condition is satisfied as close as $R \sim 20$ au in the $\rm St = 0.1$ case.

Figure \ref{fig:mfrag} shows the initial fragment masses for the two cases. \cite{Xu24} showed that clump masses forming in their simulations are distributed log-normally with the $1 \sigma$ spread of $\sigma(\log M_{\rm frag}) = 0.57$, and this spread is shown using the band around our dust growth case. We observe from the figure that the minimum $M_{\rm frag}$ is about a factor of two lower when dust growth up to $\text{St}_\text{max} = 0.1$ is allowed. While in itself this is a relatively small effect,  the minimum fragment mass $M_\text{min} \approx 5 M_J$ occurs at $R_\text{min} \approx 30$ au, which is much smaller than for the ISM-like dust, when the minimum is at $R_{\rm min}\sim 60$ au. Since protoplanetary discs are observed to extend to $20-30$ au on average \citep{Trapman23, GuerraAlvarado25}, the possibility of planet-mass fragments at small $R$ is especially important\footnote{although note that the epoch of disc fragmentation we are most concerned here could be limited to much earlier disc ages, $t\lesssim 0.5$~Myr, while material infall from the dusty envelope is ongoing \citep[e.g.,][]{VB06,VB15}, and these discs could be more extended than older ALMA discs.}.

In the case of $a_{\rm max} = 1$ cm, although the minimum is at $R_\text{min} \approx 40$ au, which is larger than the average disc extent stated above, the formation of planet-mass clumps at $R = 20-30$ au remains possible. For $v_{\rm frag} = 10$ m s$^{-1}$, planet-mass clumps are less likely to form at that radius range, but not impossible due to the spread in fragment masses as shown in fig. \ref{fig:mfrag} for the $\text{St}_\text{max} = 0.1$ case. 

\bref{Since the critical $\beta$ is directly correlated to our choice of $\alpha_c$, we have explored the effects of varying $\alpha_c$ by a factor of $2$ higher and lower, which correspond to $\beta \approx 3$ and $\beta \approx 15$ respectively. It is found that $\dot M$ increases by a factor of $\sim 2$ with higher $\alpha_c$, and vice versa. However, $T$, $\Sigma$, and $a_{\rm max}$ show basically no change for $R > 50$. For $R < 50$, the quantities vary by no more than $10\%$, increasing and decreasing with $\alpha_c$. The profile also shifts radially by a few au, outwards for higher $\alpha_c$ and vice versa. The resulting $M_{\rm frag}$ also varies by $\sim 10\%$ with higher $\alpha_c$ and vice versa. To summarise, an increase in the critical $\beta$ can lower the critical accretion rate, though it only weakly facilitates fragmentation otherwise.}

We have also explored the effects of stellar mass, disc metallicity, and irradiation on fragmentation conditions and fragment masses (available as supplementary material). In general, fragmentation is favoured in low opacity and irradiation environments (e.g. lower mass stars, irradiation shielding, metal-poor systems), which is consistent with findings from simulations \citep{MeruBate10b}.

\section{Summary and Discussion}\label{sec:discuss}

We investigated the effects of dust growth on disc fragmentation conditions. We find that dust growth beyond $a_{\rm max}\sim 1 $~mm reduces the disc opacity sufficiently to allow fragmentation to occur in colder, less massive, discs, and at smaller radii, i.e., $R\sim 30$ versus $R\sim 60$~au (cf. Fig. \ref{fig:critdisc}) for the standard ISM dust opacity. Dust growth also reduces the initial fragment mass (Fig. \ref{fig:mfrag}). As a result, disc fragmentation on planetary mass clumps, $M_{\rm frag} \leq 10 \mj$, may be more likely than usually concluded, but we note that our analytical calculations rely on a number of assumptions;  numerical simulations are needed to explore the problem further.

We assumed that dust properties depend only on the radial coordinate, and that the metallicity (which encompasses both solids and vaporised materials) is constant throughout the disc volume. Numerical simulations show large inhomogeneities in the distribution of both solids \citep[e.g.,][]{BoleyDurisen10, ForganRice11,BoothClarke16,Longarini23b,Birnstiel24} and volatile materials \citep[][]{Molyrova_21} in self-gravitating discs. These effects are multi-faceted and future dedicated numerical studies are required. For example, vertical dust settling to the disc midplane may reduce disc column-averaged dust opacity even further compared to our estimations. However, preferential dust accumulation in spiral density arms may increase the opacity there, making these regions less likely to fragment onto clumps.

Furthermore, simulations show that after forming, the fragments evolve significantly within the disc. The fragments gain mass by merging \citep{HallCEtal17} or by accreting material from the disc \citep{ZhuEtal12a,StamatellosInutsuka18}, yet they can lose mass due to tidal effects \citep[e.g.][]{Nayakshin17a, Vorobyov-Elbakyan-18}. Therefore, simulations are required to establish the eventual outcome of disc fragmentation. We note in passing that grain growth will also affect gas accretion onto the fragments since gas can only accrete onto a planet if the cooling rate is sufficiently high \citep{PollackEtal96,AyliffeBate09}.

\section{Acknowledgement}

We thank Dimitris Stamatellos for helpful comments on the draft. HL acknowledges STFC PhD studentship funding. RAB acknowledges support from the Royal Society through a University Research Fellowship and Enhanced Expenses Award.

\section{Data availability}

The opacity function and parameter table are available online\footnote{https://github.com/hanslee6/dust\_growth\_opacity}.

\newpage


\bibliographystyle{mnras}
\bibliography{nayakshin, hanslee, hanslee2, RAB}

\bsp	
\label{lastpage}
\end{document}


\maketitle

In the following figures, we consider the $\text{St}_{\rm max} = 0.1$ case described in \S 3, and all modifications are stated explicitly.

\begin{figure}
\centering
\includegraphics[width=0.7\columnwidth]{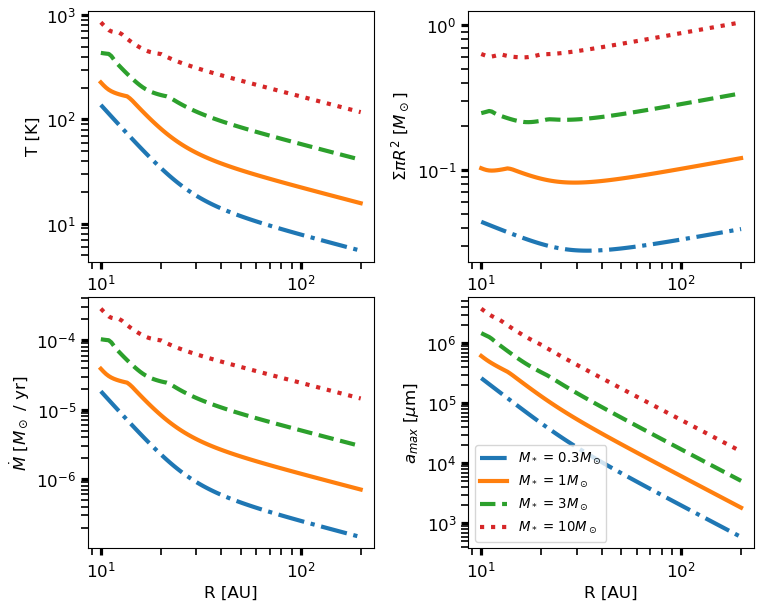}
\caption{$T_{\rm crit}$, $\Sigma_{\rm crit}$, $\dot M_{\rm crit}$, and $a_{\rm max}$ plotted against $R$ for different values of $M_*$.}
\label{fig:mstar}
\end{figure}

\begin{figure}
\centering
\includegraphics[width=0.7\columnwidth]{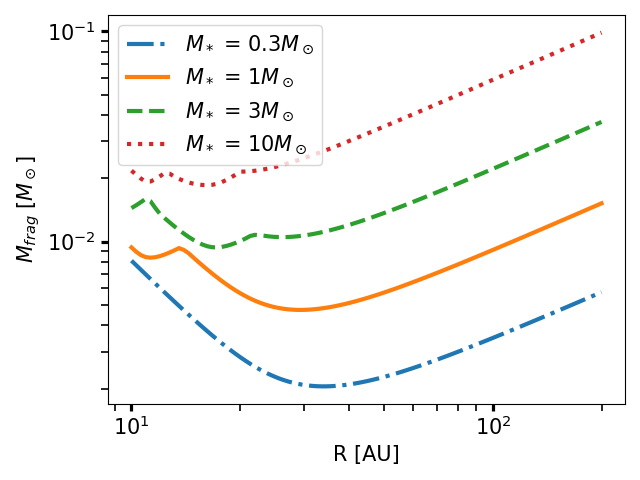}
\caption{$M_\text{frag}$ plotted against $R$ for different values of $M_*$.}
\label{fig:mstar2}
\end{figure}

Fig. \ref{fig:mstar} shows how stellar mass $M_*$ influences fragmentation conditions. Here we only consider one-star systems, and use the luminosity scaling relation for main sequence stars: $L_* \propto M_*^{3.5}$. As the irradiation temperature increases with stellar mass, according to $T_{\rm irr}^4 \propto L_* \propto M_*^{3.5}$, the disc heats up much more before reaching thermal equilibrium. As shown in \S 3, $\Sigma_{\rm crit}$ and $\dot M_{\rm crit}$ also increase with the temperature.

Fig. \ref{fig:mstar2} shows how $M_*$ influences the fragment mass. The minimum value of $M_{\rm frag}$ increases by a factor of $\sim 2$ for a factor of $3$ increase in $M_*$. This suggests that the average mass of companions increases with the mass of the central star, which is not unexpected. It should also be noted that the position of the minimum moves inward with increasing $M_*$.

\begin{figure}
\centering
\includegraphics[width=0.7\columnwidth]{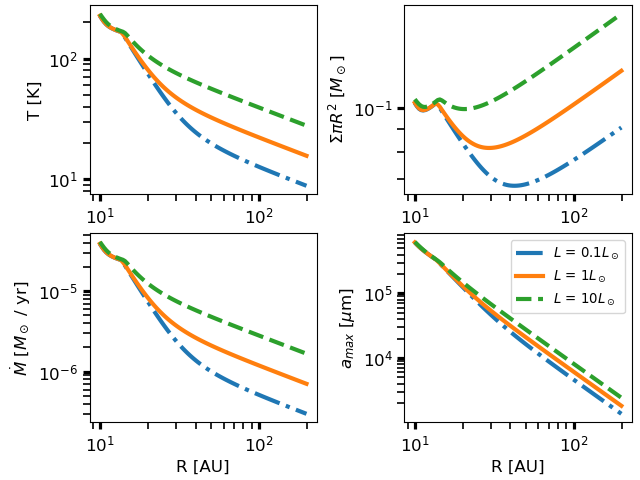}
\caption{$T_{\rm crit}$, $\Sigma_{\rm crit}$, $\dot M_{\rm crit}$, and $a_{\rm max}$ plotted against $R$ for different values of $L_*$.}
\label{fig:irr}
\end{figure}

\begin{figure}
\centering
\includegraphics[width=0.7\columnwidth]{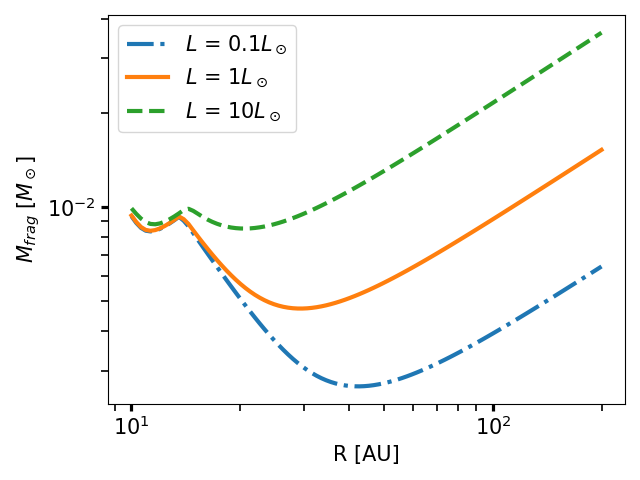}
\caption{$M_\text{frag}$ plotted against $R$ for different values of $L_*$.}
\label{fig:irr2}
\end{figure}

Fig. \ref{fig:irr} shows how the amount of stellar irradiation influences fragmentation conditions. We model this effect by multiplying $L_*$ by a fixed factor. Young stars ($\lesssim 1$ Myr) may have a higher luminosity than main sequence stars, which would increase the amount of irradiation incident on the disc. Alternatively, geometrical effects and dust above the disc may be able to block irradiation and therefore reduce irradiation heating. We observe that lower irradiation levels reduces $T_{\rm crit}$, $\Sigma_{\rm crit}$, and $\dot M_{\rm crit}$. The general trend is the same as the one observed in fig. \ref{fig:mstar}, but we note that the effect is insignificant at $R \sim 10$ au, and becomes stronger as $R$ increases.

Fig. \ref{fig:irr2} shows how stellar irradiation influences the fragment mass. We confirm that as $L_*$ decreases, $M_{\rm frag}$ shows a greater decrease at large $R$, and the minimum $M_{\rm frag}$ is attained at a larger $R$. These results show that stellar activity and shielding could strongly influence disc fragmentation, and future studies should explore this effect in more realistic scenarios.

\begin{figure}
\centering
\includegraphics[width=0.7\columnwidth]{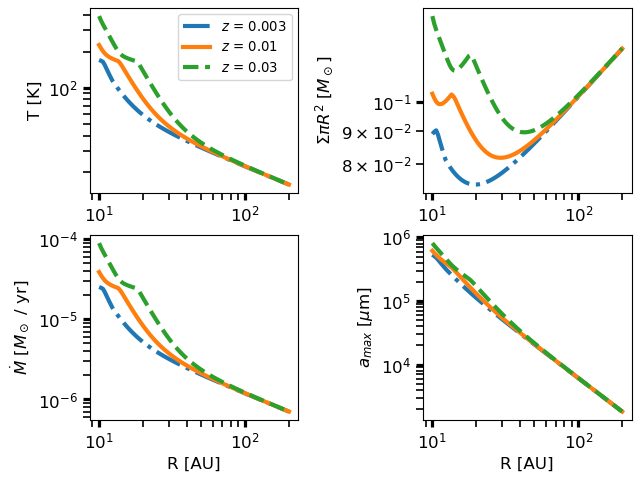}
\caption{$T_{\rm crit}$, $\Sigma_{\rm crit}$, $\dot M_{\rm crit}$, and $a_{\rm max}$ plotted against $R$ for different values of $z$.}
\label{fig:metal}
\end{figure}

\begin{figure}
\centering
\includegraphics[width=0.7\columnwidth]{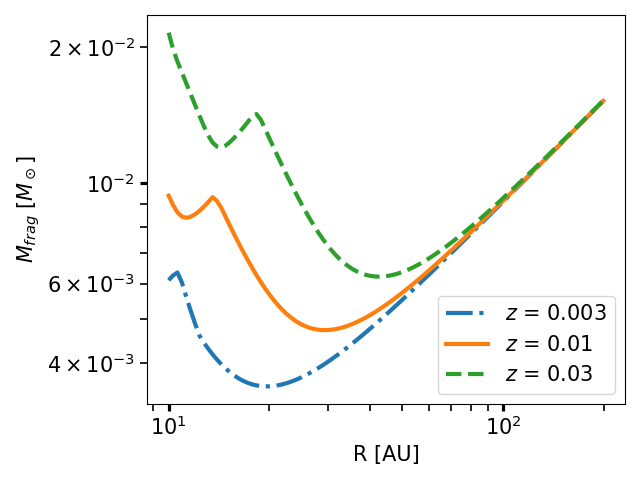}
\caption{$M_\text{frag}$ plotted against $R$ for different values of $z$.}
\label{fig:metal2}
\end{figure}

Fig. \ref{fig:metal} shows how the metallicity $z$ influences fragmentation conditions. Since dust dominates the opacity at this temperature range, the opacity increases linearly with $z$. This leads to slight increases in $T_{\rm crit}$, $\Sigma_{\rm crit}$, and $\dot M_{\rm crit}$. We note that the effect is stronger for lower $R$, but becomes very small beyond $R \sim 60$ au as the curves converge.

Fig. \ref{fig:metal2} shows that the minimum $M_{\rm frag}$ increases with $z$. However, a smaller minimum $M_{\rm frag}$ is attained at a smaller $R$, which is the opposite of our findings in varying $M_*$ and $L_*$. These suggest that metal-poor discs fragment more easily and form less massive clumps, but the dependence is not strong.